\documentclass[runningheads,citeauthoryear]{apinv}
\usepackage{epsfig,cite,graphics}
\usepackage{marvosym} 

\usepackage[utf8]{inputenc}

\begin{document}

\title{A Homogeneous Determination of the Interstellar Extinction Law and Metallicity for 105 Galactic Open Clusters}
\author{Tahereh Ramezani}
\authorrunning{T.Ramezani}
\tocauthor{Tahereh Ramezani} 
\institute{Department of Theoretical Physics and Astrophysics, Masaryk University, \\ Kotl\'a\v{r}sk\'a 2, 611\,37 Brno, Czechia   \newline
	\email{ramezani@physics.muni.cz}    }
\papertype{Submitted on 25.06.2026}	

\maketitle

\begin{abstract}
This work presents an investigation of the interstellar extinction law toward 105 Galactic open clusters using ultraviolet–optical photometry. The photometric dataset and reddening determinations were previously introduced in Ramezani et al., (2025); hereafter Paper I, where homogeneous ultraviolet photometry and colour–colour diagrams were used to derive reddening values for the cluster sample.
\\
In the present study, we extend that analysis by determining extinction law and metallicity parameters for the same clusters. Ground-based U-band observations obtained with the 2.15-m CASLEO telescope and the 1.54-m Danish telescope are combined with Gaia Data Release 3 photometry to construct ultraviolet–optical colour indices. Using extinction-ratio combinations and the Fitzpatrick, (1999) extinction model, we derive mean total-to-selective extinction law values ($R_V$) for all clusters.
\\
The resulting $R_V$ values span approximately 2.5–4.5, with a mean value of $R_V$ = 3.53, indicating significant spatial variations of dust properties across the Galactic disk and deviations from the canonical diffuse interstellar medium value $R_V$ = 3.1.
\\
Afterward, by fitting the isochrone to the Colour-Magnitude Diagrams, we estimated the metallicity values of 105 observed clusters.
\\
This work provides the first homogeneous catalogue of extinction law measurements based on combined ground-based ultraviolet and Gaia photometry for a large open-cluster sample, establishing an important reference for studies of Galactic dust structure and stellar parameter corrections.
\end{abstract}

\section*{Introduction}
Interstellar extinction is one of the principal sources of uncertainty in determining stellar parameters and Galactic structure. The wavelength dependence of extinction, commonly parameterized by the total-to-selective extinction ratio $R_V$, reflects the physical properties of interstellar dust grains such as size distribution and composition. Although a canonical value of $R_V \approx 3.1$ is often adopted for the diffuse interstellar medium, numerous studies have demonstrated that the extinction law varies significantly across different Galactic environments.
\\
Open clusters provide ideal laboratories for investigating variations in the extinction law because their member stars share common distances and ages. In particular, ultraviolet (UV) observations are highly sensitive to reddening effects, since extinction increases strongly toward shorter wavelengths. Combining ground-based U-band photometry with precise optical measurements from Gaia Data Release 3 Gaia Collaboration, (2023) enables extinction behaviour to be investigated across a broad wavelength baseline. This combination significantly improves sensitivity to deviations from the standard Galactic extinction law and allows reliable characterization of dust properties along different lines of sight.
\\
This study is a direct continuation of our previous work, \textit{Ultraviolet photometry and reddening estimation of 105 Galactic open clusters} Ramezani et al., (2025), hereafter Paper~I. In Paper~I, homogeneous ultraviolet photometry was presented, and colour--colour diagrams were used to derive reddening values for 105 Galactic open clusters based on combined U-band and Gaia photometry.
\\
The present paper does not repeat the observational analysis or reddening determination. Instead, it uses the homogeneous reddening catalogue established in Paper~I to address a different astrophysical problem; the determination of extinction law variations across the Galactic disk. By analysing colour-excess ratios involving the U, BP, G, and RP passbands, we derive total-to-selective extinction values ($R_V$) using the extinction model of (Fitzpatrick, (1999)).
\\
Although Paper I established the uniform ultraviolet photometric framework for these 105 clusters, a systematic determination of the total-to-selective extinction parameter ($R_V$) and its spatial variations across the Galactic plane remained unaddressed. This work builds directly on that foundation to provide a unified map of interstellar dust properties and cluster metallicities.

\section*{1. Adopted Observational Data}
The photometric dataset used in this study was presented in detail in Paper I, where ultraviolet observations and reddening determinations were carried out for 105 Galactic open clusters.
\\
Briefly, the clusters were observed in the ground-based U band using the 2.15-m telescope at CASLEO (Argentina) and the 1.54-m Danish Telescope at La Silla (Chile). Standard data reduction procedures, including bias subtraction, flat-field correction, and PSF photometry, were performed using IRAF routines as described in Paper I.
\\
Observed sources were cross-matched with Gaia Data Release 3 photometry (G, BP, and RP bands) using the Match pipeline based on the FOCAS algorithm (Valdes et al., (1995)). Photometric calibration onto the Gaia system was achieved through a multivariate linear transformation following (Viscasillas et al., (2024)).
\\
In the present work, we use the homogeneous photometric catalogue and reddening measurements derived in Paper I as input data to determine the extinction law parameters.

\section*{2. Ultraviolet–Optical Colour Analysis}

Colour-colour diagrams constructed from U-band and Gaia DR3 photometry were used in Paper I to determine reddening values for all clusters. The diagrams combine ultraviolet sensitivity to hot stars with the precise Gaia optical photometric system, allowing reliable identification of cluster sequences and reddening estimation.
\\
Padova isochrones Bressan et al., (2012) were fitted to cluster members identified from the catalogue of (Hunt \& Reffert, (2023)). The resulting colour excesses derived in Paper I form the basis of the present analysis.
\\
In this study, the previously determined extinction ratios are not re-derived; instead, they are used to investigate the interstellar extinction law and to estimate $R_V$ values for the full cluster sample.

\section*{3. Estimation of $R_V$ Values}

\begin{figure}[htb]
    \centering
    \includegraphics[width = \columnwidth]
    {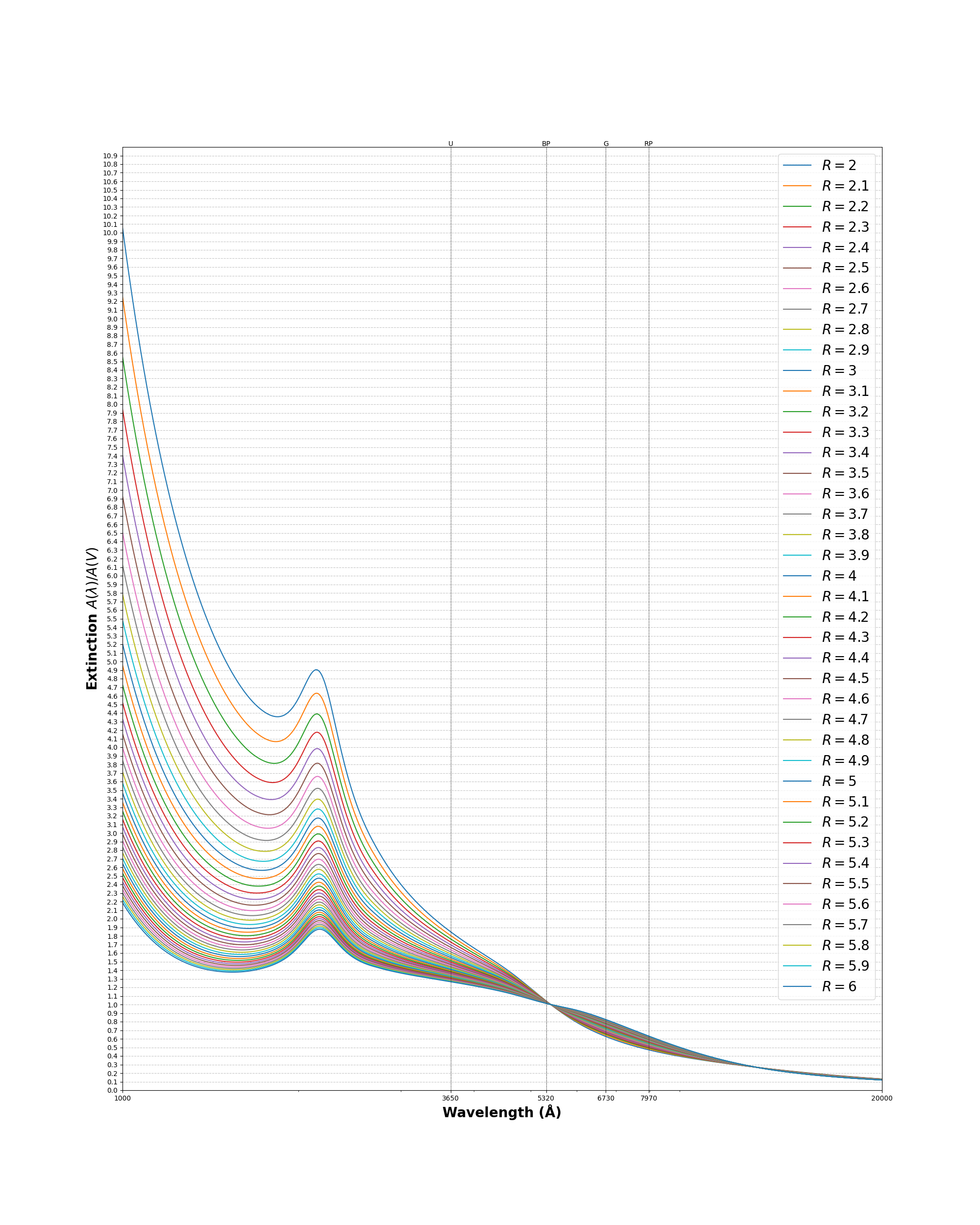}
    \caption{Wavelength (\AA) vs. extinction ($A_\lambda / A_V$).}
    \label{Fig.1}
\end{figure}

\begin{figure}[htb]
    \centering
    \includegraphics[width = \columnwidth]
    {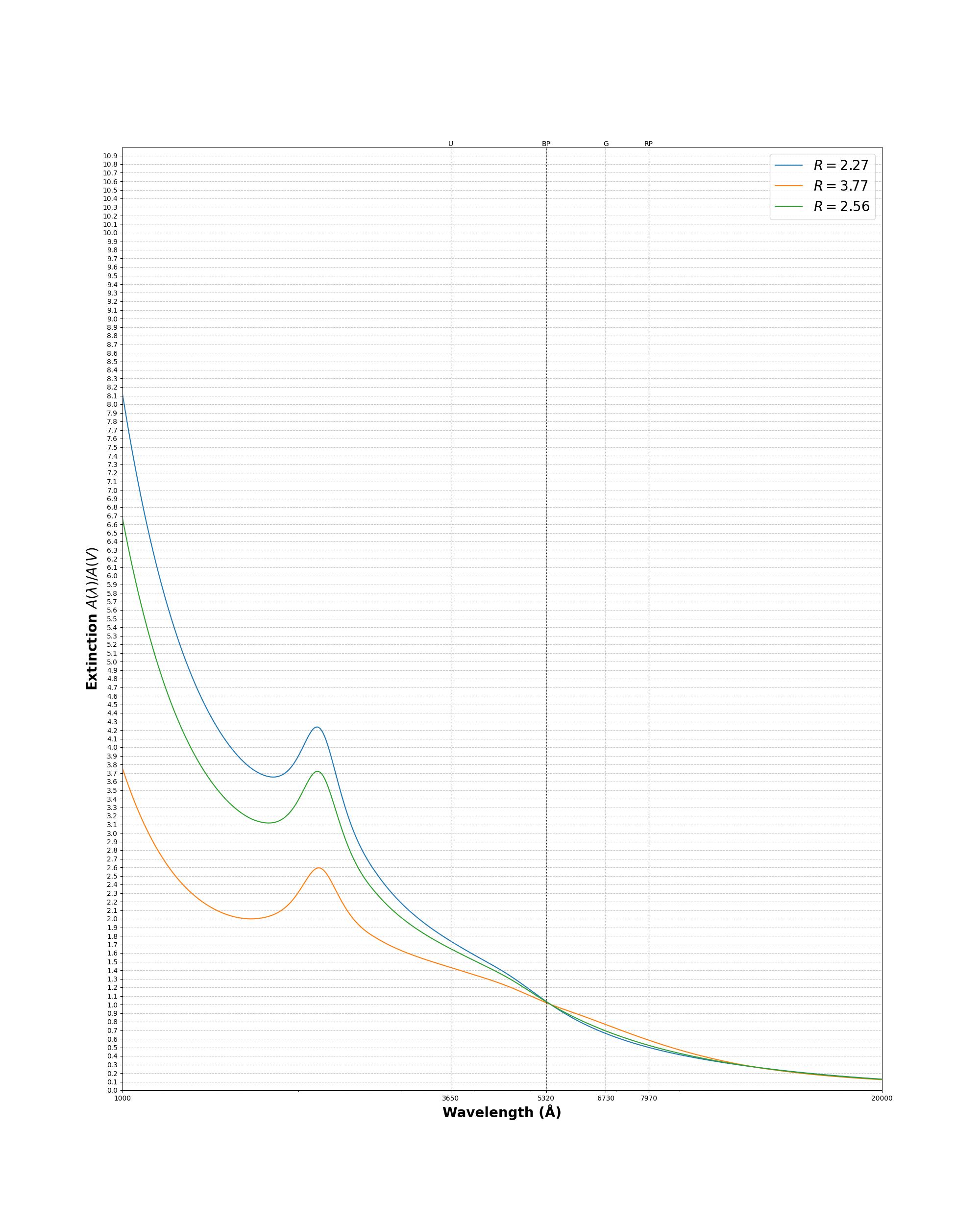}
    \caption{Wavelength (\AA) vs. extinction ($A_\lambda / A_V$) for NGC 2225 with $R_V$ = 2.27, 3.77, and 2.56.}
    \label{Fig.2}
\end{figure}

The wavelength dependence of interstellar extinction is commonly described by the total-to-selective extinction parameter $R_V = A_V / E(B-V)$, which is sensitive to the physical properties of interstellar dust grains. Larger $R_V$ values generally correspond to environments containing larger dust grains or denser interstellar regions, while lower values indicate diffuse media dominated by smaller grains.
\\
In this work, $R_V$ values are derived from ultraviolet--optical colour-excess ratios obtained from the homogeneous reddening measurements presented in Paper~I. The conversion from observed colour excesses to extinction law parameters is performed using the Fitzpatrick, (1999) extinction model.
\\
Fig.~\ref{Fig.1} shows theoretical extinction curves expressed as $A_\lambda/A_V$ as a function of wavelength for $2 \leq R_V \leq 6$, generated using the implementation of (Barbary, (2021)). The effective wavelengths of the U, BP, G, and RP filters are indicated, allowing direct comparison between observed colour-excess ratios and theoretical extinction behaviour.
\\
For each cluster, extinction ratios constructed from ultraviolet and Gaia passbands were compared with the theoretical extinction curves. Only curves consistent with the measured colour-excess ratios were considered, providing an estimate of the corresponding $R_V$ value.
\\
As an example, Fig.~\ref{Fig.2} illustrates the procedure for the open cluster NGC~2225. $R_V$ estimates are obtained from several colour-excess combinations:

\begin{equation}
\frac{E(U-RP)}{E(BP-RP)} = 2.27,
\end{equation}

\begin{equation}
\frac{E(U-G)}{E(G-RP)} = 3.77,
\end{equation}

\begin{equation}
\frac{E(U-G)}{E(BP-G)} = 2.56.
\end{equation}
\\
The final cluster value is taken as the mean of the individual estimates,

\begin{equation}
 R_V  = 2.86,
\end{equation}
\\
while the uncertainty corresponds to the mean propagated uncertainty derived from the individual extinction ratios.
\\
This procedure is applied uniformly to all 105 clusters, yielding a homogeneous catalogue of $R_V$ values. The resulting mean extinction parameters and their uncertainties are listed in Table 1. In its complete form, this table is only available at the CDS or upon request.

\begin{table*}[htb]

   \centering
   \label{Table 1}   
   \caption{Mean $R_V$ values and mean uncertainties for all 105 observed clusters.}
   \vspace{0.5cm}

   \begin{tabular}{llccc}
   \hline\hline
   
Cluster & Mean $R_V$ value & Mean uncertainty \\
   \hline
Alessi  17 & 4.09 & 0.35 \\
Alessi  60 & 4.39 & 0.15 \\
Berkeley  33 & 4.02 & 0.14 \\
BH  72 & 4.02 & 0.14 \\
BH  84 & 3.88 & 0.22 \\
BH  87 & 3.67 & 0.15 \\
BH  111 & 4.22 & 0.34 \\
BH  132 & 2.66 & 0.35 \\
BH  140 & 3.38 & 0.20 \\
CWNU  95 & 2.84 & 0.39 \\
CWNU  1733 & 2.94 & 0.47 \\
Czernik  29 & 3.77 & 0.16 \\
   \hline
   \hline  

   \end{tabular}

\end{table*}

\clearpage

\begin{figure}[htb]
    \centering
    \includegraphics[width = \columnwidth]
    {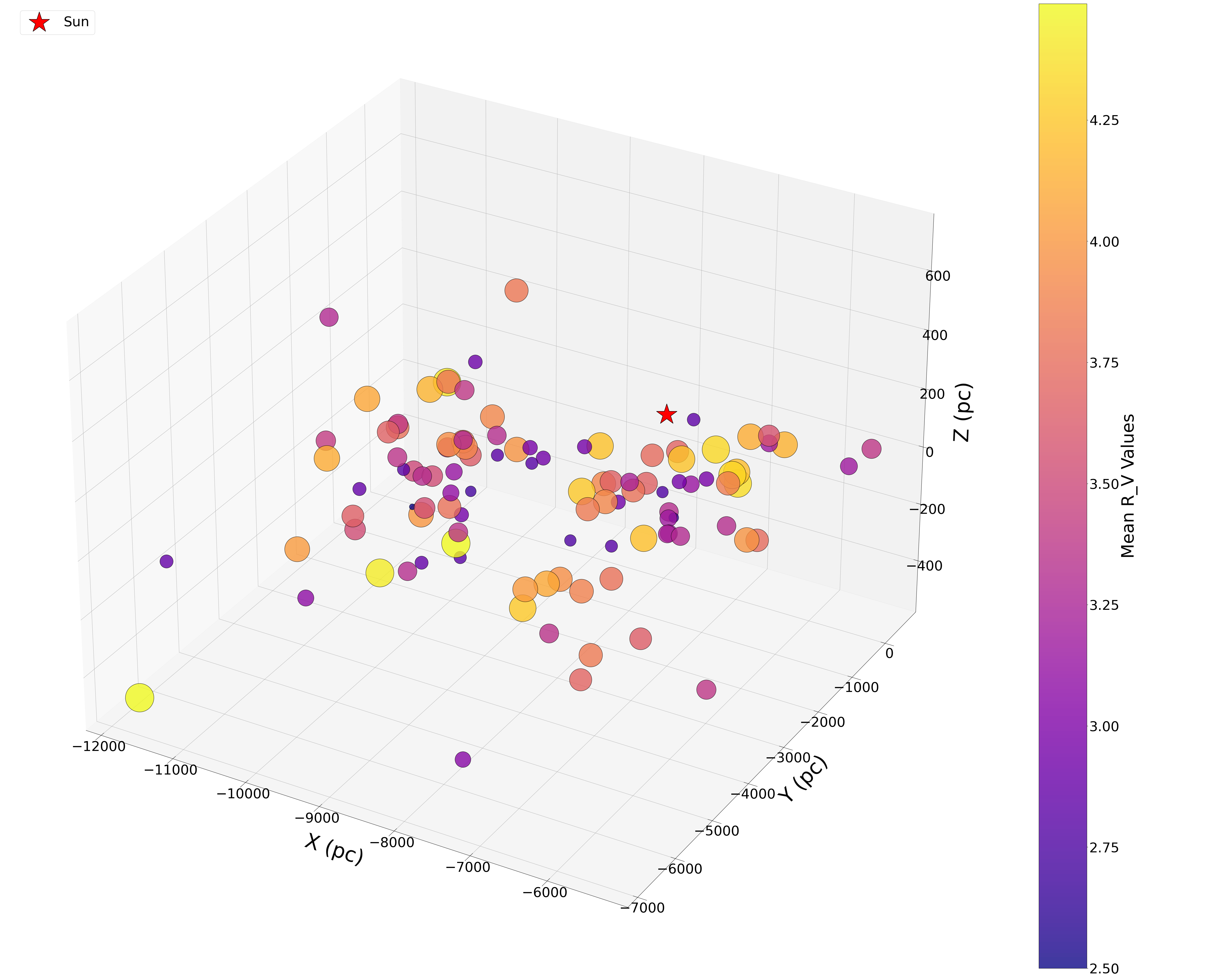}
    \caption{3D plot of the 105 observed clusters with their mean $R_V$ values in the galactic coordinate system [X, Y, Z].}
    \label{Fig.3}
\end{figure}

To investigate possible spatial trends of the extinction law, we examined the Galactic distribution of the cluster sample. Fig. \ref{Fig.3} presents the three-dimensional Galactic positions ([X, Y, Z]) adopted from (Hunt \& Reffert, (2023)). The coordinate system is defined such that the X-axis points toward the Galactic centre, the Y-axis follows the direction of Galactic rotation, and the Z-axis is perpendicular to the Galactic plane.
\\
The symbol size represents the derived mean $R_V$ value for each cluster. The vertical distribution extends from approximately $Z=-495$ pc to $Z=+702$ pc, sampling both thin-disk and transitional disk environments. The absence of a single preferred $R_V$ value across the sampled volume indicates that extinction law variations are not confined to a specific Galactic location, but instead reflect local interstellar medium conditions.

\vspace{0.75cm}

\clearpage

\begin{figure}[htb]
    \centering
    \includegraphics[width= \columnwidth]
    {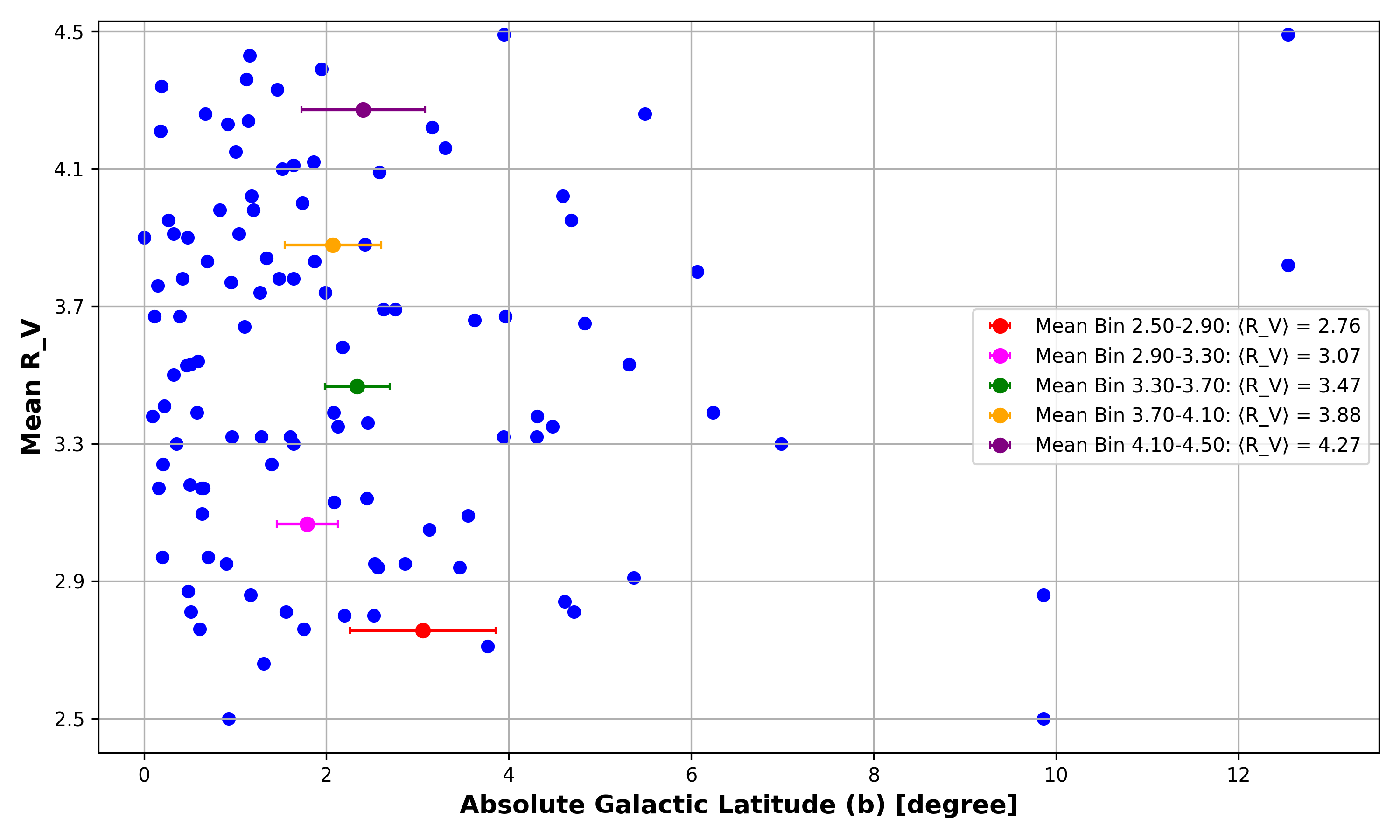}
    \caption{Galactic latitude vs. Mean $R_V$. Values of Galactic latitude are from (Hunt \& Reffert, (2023)).}
    \label{Fig.4}
\end{figure}

Fig. \ref{Fig.4} shows the dependence of the mean $R_V$ values on Galactic latitude. Larger $R_V$ values are predominantly concentrated near the Galactic plane, consistent with environments containing denser interstellar material and larger dust grains. The increased dispersion at higher latitudes suggests the presence of localized dust structures and supports the interpretation that extinction law variations arise from environmental processing rather than a universal Galactic trend.

To validate the robustness of our method, we applied the same procedure to several clusters previously analysed in independent studies Akkaya Oralhan et al., (2024) and (Legnardi et al., (2023)). Using Gaia DR3 photometry and synthetic U-band magnitudes, cluster members were re-analysed following the methodology described above. The resulting $R_V$ values show good agreement with published literature measurements, confirming the reliability and reproducibility of our extinction law determination technique. The derived extinction ratios and comparison results are summarized in Table \ref{Table 2}.

\begin{table}[htb]

   \centering
   
    \caption{The columns denote: (1) Cluster name. (2) Right ascension (J2000; Gaia DR3). (3) Declination (J2000; GaiaDR3). (4) Our estimated E(U-BP)/E(BP-RP). (5) Our estimated E(U-G)/E(BP-RP). (6) Our estimated E(U-RP)/E(BP-RP). (7) Our estimated E(U-G)/E(G-RP). (8) Our estimated E(U-G)/E(BP-G). (9) Our estimated mean $R_V$ values. (10) Our estimated mean uncertainties. (11) Mean $R_V$ values of literature. (12) Mean uncertainties of literature.}
   
   \vspace{0.5cm}

   \resizebox{\textwidth}{!}{ 
   \Large
   \begin{tabular}{llcccccccccccc}
   \hline\hline
   (1) & (2) & (3) & (4) & (5) & (6) & (7) & (8) & (9) & (10) & (11) & (12) \\
   \hline
   \large NGC 1513 &\large 062.47155193 &\large +49.50155413 &\large 1.75 &\large 2.08 &\large 2.85 &\large 3.08 &\large 5.50 &\large 2.78 &\large 0.05  &\large 2.85 &\large 0.05 \\
   \large NGC 6093 &\large 244.23534723 &\large -22.97018516 &\large 0.89 &\large  1.20 &\large 2.03 &\large 2.21 &\large 4.20 &\large 3.99 &\large 0.14  &\large 3.93 &\large 0.18 \\
   \large NGC 6254 &\large 254.24375035 &\large -04.10527590 &\large 0.82 &\large 1.70 &\large 2.20 &\large 2.86 &\large 2.80 &\large 3.42 &\large 0.04  &\large 3.35 &\large 0.27 \\
   \large NGC 6656 &\large 279.10570738 &\large -23.95327177 &\large 0.74 &\large 1.21 &\large 2.00 &\large 1.71 &\large 3.81 &\large 2.57 &\large 0.05  &\large 2.53 &\large 0.11 \\
   \large NGC 6779 &\large 289.13087838 &\large +30.17744477 &\large 0.54 &\large 0.74 &\large 1.42 &\large 1.47 &\large 2.58 &\large 2.91 &\large 0.05  &\large 2.87 &\large 0.19 \\
   \hline
   \hline  

   \label{Table 2}
    \end{tabular}}

\end{table}

\section*{4. Estimation Of Metallicity}

To estimate the metallicity of each observed cluster, we plot the Colour-Magnitude Diagram (CMD). To plot the CMD, we need age, distance, reddening, and metallicity. We have age and distance from Hunt \& Reffert, (2023), and reddening from our analysis. We use the PARSEC isochrone Bressan et al., (2012), the input metallicities range from 0.001 to 0.03 with a step of 0.001.   
\\
Then we plot the CMD of (BP-RP) [mag] vs. G [mag] of the members, with fixed age, distance, reddening, and metallicity as a free parameter. 
\\
We apply a program that takes the PARSEC isochrones as a model, which defines theoretical isochrones for a given metallicity Z. (BP-RP) and G as data, which defines observed cluster stars (pink points). 
For every observed star, it computes the distance in the CMD to the closest point on the isochrone:

\begin{equation}
    \chi^2 = \sum_{i} d_i^2   
    \label{equation5}
\end{equation}

\begin{figure}[htb]
    \centering
    \includegraphics[width = \columnwidth]
    {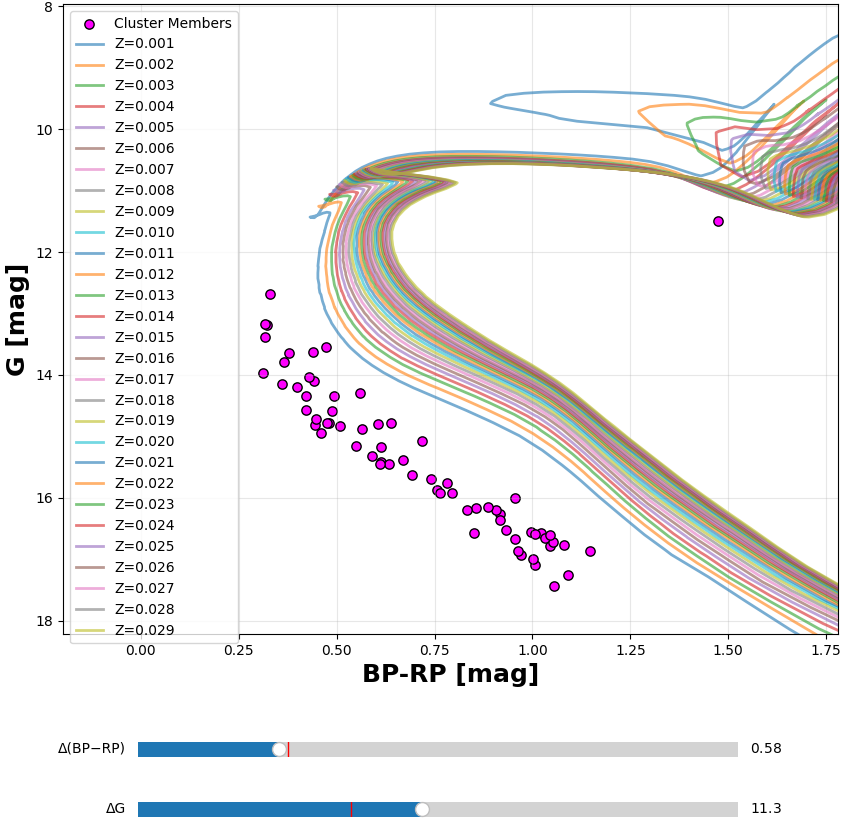}
    \caption{(BP-RP) [mag] vs. G [mag] with metallicities from 0.001 to 0.03 for Alessi 60.}
    \label{Fig.5}
\end{figure}

\begin{figure}[htb]
    \centering
    \includegraphics[width = \columnwidth]
    {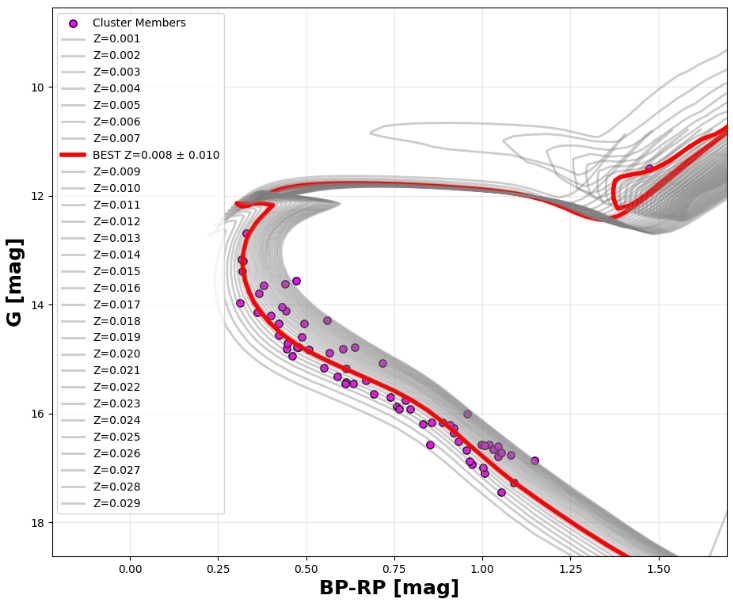}
    \caption{(BP-RP) [mag] vs. G [mag] with the best metallicity and uncertainty for Alessi 60.}
    \label{Fig.6}
\end{figure}

In equation \ref{equation5}, $d_i$ is the distance between star $i$ and the isochrone, and $\sum_{i}$ is the sum that runs over all stars. Therefore, stars very close to the isochrone have small distances, then have small $\chi^2$, which gives a better fit; and stars far away have large distances, then have large $\chi^2$. 
\\
We used the sliders for $\Delta$(BP-RP), which shifts colour (reddening), and $\Delta$G, which shifts magnitude (distance modulus). We visually place the isochrones where the cluster sequence lies.
\\
For each metallicity, first, it takes the corresponding isochrone, second, applies our chosen shifts ($\Delta$ colour, and $\Delta$ magnitude), third, compares all stars with that isochrone, and finally computes $\chi^2$.
\\
The program then selects the best Z, which is the metallicity with the smallest $\chi^2$, because it corresponds to the isochrone closest to the observed cluster sequence. The best metallicity is the one that minimizes the total distance between the stars and the isochrone.
\\
The uncertainty in the metallicity determination is estimated by evaluating the isochrone fit's sensitivity to changes in the adopted metal abundance. After the metallicity that provides the best visual and statistical agreement between the observed cluster sequence and the theoretical isochrones is identified, neighboring metallicity models are also examined. Metallicities slightly higher than or lower than the optimal value are considered acceptable as long as the isochrone continues to reproduce the main-sequence location of the cluster within the observational scatter of the data. The range of metallicities that still produce comparably good fits defines the allowed interval of solutions. The metallicity uncertainty is then derived from the width of this interval, which represents the level of variation in the metal abundance that does not significantly degrade the agreement between observations and models.
\\
Fig. \ref{Fig.5} shows the CMD for Alessi 60, with metallicities ranging from 0.001 to 0.03 before applying the method to determine the best metallicity. 
\\
Fig. \ref{Fig.6} shows the CMD for Alessi 60, with the best metallicity and its uncertainty.
\\
Metallicity values and their uncertainties are given in Table \ref{Table 3}. In its complete form, this table is only available at the CDS or upon request.

\vspace{1 cm}

\begin{table}[htb]

   \centering
   
   \caption{Metallicity values and their uncertainties for all 105 observed clusters. BH 140 doesn't have LogAge (year) in Hunt \& Reffert, (2023), so we couldn't estimate its metallicity.}
   
   \vspace{0.5cm}

   \begin{tabular}{llccc}
   \hline\hline
   
Cluster & Metallicity & uncertainty \\
   \hline
Alessi  17 & 0.005 & 0.014 \\
Alessi  60 & 0.008 & 0.01 \\
Berkeley  33 & 0.005 & 0.006 \\
BH  72 & 0.029 & 0.004 \\
BH  84 & 0.019 & 0.011 \\
BH  87 & 0.006 & 0.009 \\
BH  111 & 0.005 & 0.013 \\
BH  132 & 0.019 & 0.011 \\
BH  140 &  &  \\
CWNU  95 & 0.008 & 0.014 \\
CWNU  1733 & 0.01 & 0.014 \\
Czernik  29 & 0.008 & 0.010 \\
   \hline
   \hline  
    \label{Table 3}
    \end{tabular}
    
\end{table}

To test our metallicity method, we used the metallicity of 63 standard clusters, which are in Li et al., (2025), Cavallo et al., (2024), Alfonso et al., (2024), and (Netopil et al., (2022)). We follow the same process for the standard clusters and determine their metallicity using our method. The metallicities reported in the catalogues are in [Fe/H]. We convert [Fe/H] into the heavy-element mass fraction $Z$ using equation (Mowlavi et al., (2012)):

\begin{equation}
Z_{\mathrm{init}} =
\frac{1-Y_{\mathrm{init}}}
     {1+\frac{\Delta Y}{\Delta Z}
      +\frac{1}{(Z/X)_{\odot}}
       10^{-[\mathrm{Fe}/\mathrm{H}]_{\mathrm{init}}}}
       \label{Equation6}
\end{equation}

In equation \ref{Equation6}, the adopted parameters are 
$Y_{\rm init}=0.2485 + (\Delta Y/\Delta Z)Z_{\rm init}$,
$\Delta Y/\Delta Z = 2$, and
$(Z/X)_\odot = 0.02298$.
The resulting $Z$ values were used in the subsequent isochrone fitting and metallicity analysis.
\\
In Table \ref{Table 4}, we represent the min, max, median, and mean values of metallicity of Li et al., (2025), Cavallo et al., (2024), Alfonso et al., (2024), and Netopil et al., (2022) compared to our values of metallicity and uncertainty.

\vspace{0.5cm}

\begin{table}[htb]


\caption{Min, Max, Median, and Mean values of Li et al., (2025), Cavallo et al., (2024), Alfonso et al., (2024), and Netopil et al., (2022) in comparison with our estimated values of metallicity and their uncertainty.}
  
\vspace{0.5cm}

   \resizebox{\textwidth}{!}{ 
   \Large
   \begin{tabular}{llccccc}
   \hline\hline
     & Min & Max & Median & Mean \\
   \hline
   \large Li et al., (2025) metallicity &\large 0.003 &\large 0.019 &\large 0.010 &\large 0.11 \\
   \large Our metallicity estimation from Li et al., (2025) &\large 0.004 &\large 0.019 &\large 0.011 &\large 0.010 \\
   \large Our uncertainty estimation from Li et al., (2025) &\large 0.001 &\large 0.009 &\large 0.002 &\large 0.002 \\
   \large Cavallo et al., (2024) metallicity &\large 0.001 &\large 0.050 &\large 0.016 &\large 0.017 \\
   \large Our metallicity estimation from Cavallo et al., (2024) &\large 0.002 &\large 0.053 &\large 0.015 &\large 0.016 \\
   \large Our uncertainty estimation from Cavallo et al., (2024) &\large 0.001 &\large 0.100 &\large 0.003 &\large 0.005 \\
   \large Alfonso et al., (2024) metallicity &\large 0.009 &\large 0.023 &\large 0.013 &\large 0.015 \\
   \large Our metallicity estimation from Alfonso et al., (2024) &\large 0.007 &\large 0.022 &\large 0.013 &\large 0.013 \\
   \large Our uncertainty estimation from Alfonso et al., (2024) &\large 0.001 &\large 0.007 &\large 0.002 &\large 0.002 \\   
   \large Netopil et al., (2022) metallicity &\large 0.005 &\large 0.019 &\large 0.011 &\large 0.011 \\
   \large Our metallicity estimation from Netopil et al., (2022) &\large 0.004 &\large 0.020 &\large 0.011 &\large 0.010 \\
   \large Our uncertainty estimation from Netopil et al., (2022) &\large 0.001 &\large 0.006 &\large 0.002 &\large 0.002 \\ 
   \hline
   \hline  

   \label{Table 4}
    \end{tabular}}

\end{table}

\begin{figure}[htb]
    \centering
    \includegraphics[width = \columnwidth]
    {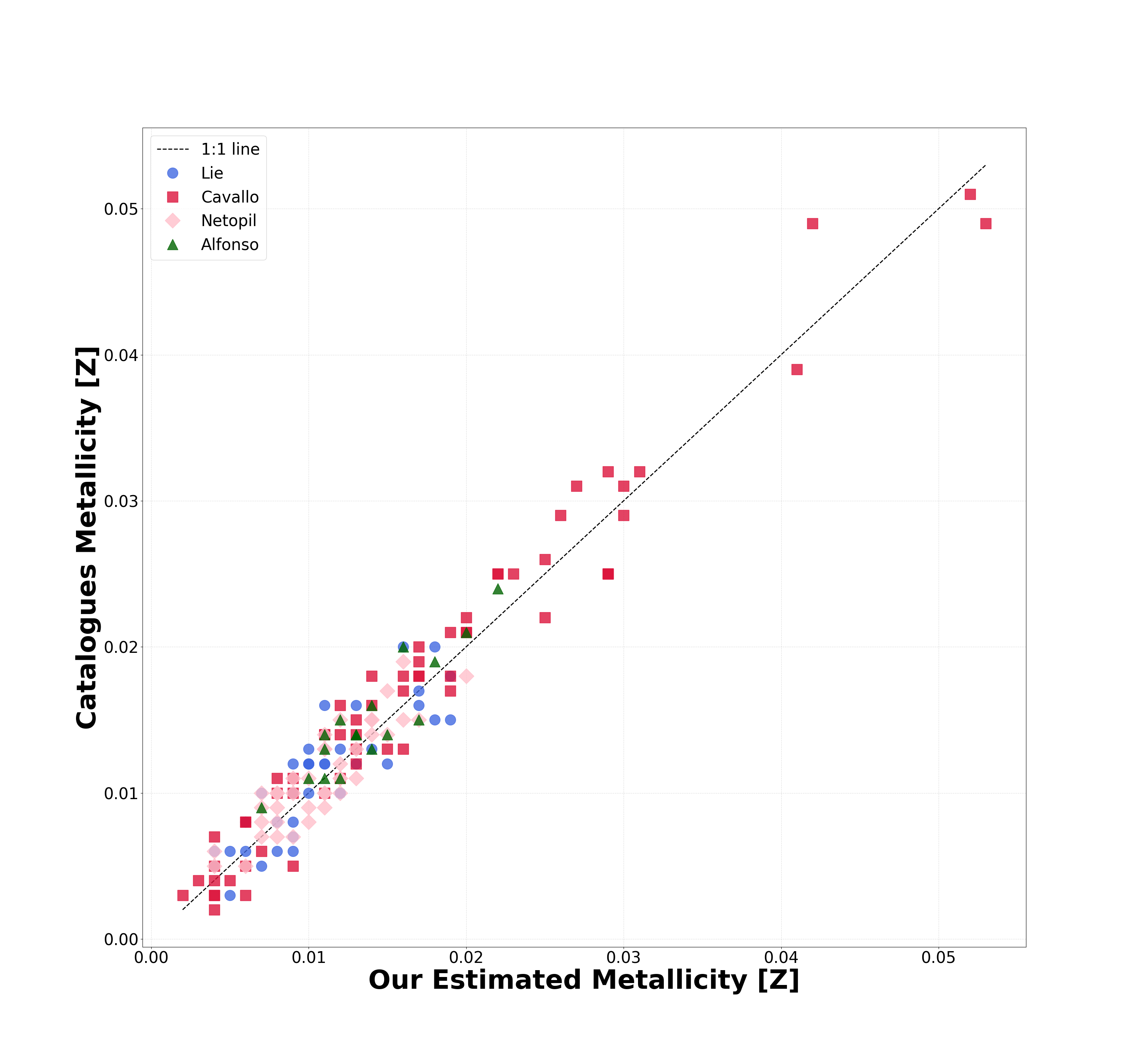}
    \caption{Comparison between our estimated metallicity vs. metallicity of Li et al., (2025), Cavallo et al., (2024), Alfonso et al., (2024), and Netopil et al., (2022) catalogues.} 
    \label{fig7}
\end{figure}

Table \ref{Table 5} represents the names of the standard clusters, their RA and DEC from Hunt \& Reffert, (2023), their metallicities in Li et al., (2025), Cavallo et al., (2024), Alfonso et al., (2024), and Netopil et al., (2022) catalogues, our estimated metallicities, and our estimated uncertainty. We see that some clusters were not considered in the catalogues. This is why there are some blank spaces in Table \ref{Table 5} and Fig. \ref{fig7}. In its complete form, Table \ref{Table 5} is only available at the CDS or upon request.

\begin{table}[htb]

   \centering

    \caption{The columns denote: (1) Cluster name. (2) Right ascension (Hunt \& Reffert, (2023)). (3) Declination (Hunt \& Reffert, (2023)). (4) Li et al., (2025) metallicity. (5) Our estimated metallicity. (6) Our uncertainty. (7) Cavallo et al., (2024) metallicity. (8) Our estimated metallicity. (9) Our uncertainty. (10) Alfonso et al., (2024) metallicity. (11) Our estimated metallicity. (12) Our uncertainty. (13) Netopil et al., (2022) metallicity. (14) Our estimated metallicity. (15) Our uncertainty.}
   
   \vspace{0.5cm}

   \resizebox{\textwidth}{!}{ 
   \Large
   \begin{tabular}{llccccccccccccccc}
   \hline\hline

   (1) & (2) & (3) & (4) & (5) & (6) & (7) & (8) & (9) & (10) & (11) & (12) & (13) & (14) & (15) \\
   \hline
   \large Alessi 13 &\large 51.992 &\large -35.773 &\large  &\large  &\large  &\large 0.022 &\large 0.020 &\large 0.004 &\large 0.24 &\large 0.022 &\large 0.005 &\large  &\large  &\large  \\
   \large Berkeley 11 &\large 65.120 &\large 44.923 &\large 0.008 &\large 0.009 &\large 0.007 &\large 0.039 &\large 0.041 &\large 0.1 &\large  &\large &\large  &\large  &\large  &\large   \\
   \large Berkeley 17 &\large 80.141 &\large 30.574 &\large 0.010 &\large 0.012 &\large 0.004 &\large 0.016 &\large 0.014 &\large 0.004 &\large  &\large  &\large  &\large 0.01 &\large 0.009 &\large 0.002  \\
   \large Berkeley 21 &\large 87.948 &\large 21.810 &\large  &\large  &\large  &\large 0.006 &\large 0.007 &\large 0.003 &\large  &\large  &\large  &\large 0.009 &\large 0.01 &\large 0.004  \\
   \large Berkeley 22 &\large 89.615 &\large 7.758 &\large  &\large  &\large  &\large 0.008 &\large 0.006 &\large 0.002 &\large  &\large  &\large  &\large 0.008 &\large 0.007 &\large 0.002  \\
   \large Berkeley 23 &\large 98.315 &\large 20.534 &\large 0.006 &\large 0.006 &\large 0.003 &\large 0.005 &\large 0.004 &\large 0.003 &\large  &\large  &\large  &\large  &\large  &\large   \\
   \large Berkeley 25 &\large 100.312 &\large -16.486 &\large  &\large  &\large  &\large 0.004 &\large 0.003 &\large 0.001 &\large  &\large  &\large  &\large 0.009 &\large 0.008 &\large 0.004  \\
   \large Berkeley 29 &\large 103.265	&\large 16.925 &\large  &\large  &\large  &\large 0.005 &\large 0.009 &\large 0.007 &\large  &\large  &\large  &\large 0.005 &\large 0.004 &\large 0.003  \\
   \large Berkeley 31 &\large 104.406 &\large 8.288 &\large 0.005 &\large 0.007 &\large 0.002 &\large 0.005 &\large 0.006 &\large 0.002 &\large  &\large  &\large  &\large 0.005 &\large 0.006 &\large 0.005 \\
   \large Berkeley 32 &\large 104.531 &\large 6.433 &\large 0.006 &\large 0.009 &\large 0.002 &\large 0.010 &\large 0.008 &\large 0.002 &\large  &\large  &\large  &\large 0.007 &\large 0.008 &\large 0.002  \\
   \large Berkeley 33 &\large 104.462 &\large -13.226 &\large 0.007 &\large 0.009 &\large 0.006 &\large 0.010 &\large 0.011 &\large 0.006 &\large  &\large  &\large  &\large 0.007 &\large 0.007 &\large 0.005  \\
   \large Berkeley 70 &\large 81.454 &\large 41.950 &\large 0.003 &\large 0.005 &\large 0.002 &\large 0.004 &\large 0.004 &\large 0.002 &\large  &\large  &\large  &\large  &\large  &\large   \\
   \hline
   \hline  

   \label{Table 5}
    \end{tabular}}

\end{table}

In Fig. \ref{fig7}, we represent the comparison of our estimated metallicity vs. the metallicity of the Li et al., (2025), Cavallo et al., (2024), Alfonso et al., (2024), and Netopil et al., (2022) catalogues. We see that there is a good agreement between our values and the catalogues' values.

\clearpage

\section*{5. Discussion}

The extinction law analysis presented in this work extends the reddening study of Ramezani et al., (2025) by investigating the physical properties of interstellar dust toward 105 Galactic open clusters. While Paper~I established homogeneous colour-excess measurements using ultraviolet and Gaia photometry, the present study interprets those measurements in terms of extinction law variations and metallicities.
\\
The derived $R_V$ values span approximately $2.5 \leq R_V \leq 4.5$, with a mean value of $R_V = 3.5$. This mean exceeds the canonical diffuse interstellar medium value of $R_V \approx 3.1$, indicating that many clusters are located in environments where dust grains are systematically larger or where dust processing has modified the grain-size distribution.
\\
Finding a mean $R_V$ of 3.53 rather than the standard diffuse ISM value of 3.1 is highly significant. Explicitly note that blindly assuming $3.1$ for these regions overestimates distance moduli and biases age estimates derived from isochrone fitting. 
\\
The significant dispersion of $R_V$ values demonstrates that the Galactic extinction law is not spatially uniform. Variations are expected because open clusters probe different Galactic environments, including diffuse regions, spiral-arm structures, and star-forming complexes. Higher $R_V$ values are generally associated with dense molecular environments where grain growth through coagulation or mantle accretion becomes efficient, whereas lower values indicate more diffuse interstellar conditions dominated by smaller grains.
\\
The use of ultraviolet–optical colour combinations provides strong sensitivity to extinction law behaviour. Ultraviolet wavelengths respond strongly to small dust grains, while Gaia optical bands trace extinction at longer wavelengths. The combination, therefore, allows robust discrimination among different extinction curves. The results confirm that ultraviolet photometry remains a powerful diagnostic of interstellar dust properties when combined with modern astrometric surveys such as Gaia DR3.
\\
An important outcome of this work is the construction of a homogeneous reddening law catalogue for a large open-cluster sample. Previous extinction studies often relied on heterogeneous datasets or limited number of clusters. The uniform methodology adopted here minimizes systematic differences between clusters and enables direct comparison of extinction properties across the Galactic disk.
\\
The observed range of $R_V$ values supports a picture in which dust properties vary continuously rather than following a single universal extinction law. Such variations have important consequences for stellar parameter determination, distance estimates, and Galactic structure studies, since incorrect extinction assumptions may introduce systematic biases in derived stellar luminosities and ages.
\\
We compare the metallicity values of 63 standard clusters reported by Li et al., (2025), Cavallo et al., (2024), Alfonso et al., (2024), and Netopil et al., (2022) with the metallicity values we obtain from our metallicity method and find agreement, confirming the accuracy of our method, which is shown in Fig. \ref{fig7}.

\clearpage

\section*{Conclusions}

We have investigated the interstellar extinction law toward 105 Galactic open clusters using ultraviolet photometry combined with Gaia DR3 data. Building upon the reddening catalogue established in Paper~I, the present work derives extinction law parameters and examines their spatial behaviour.
\\
The principal conclusions of this study are:
\\
\begin{enumerate}
\item The derived $R_V$ values range between approximately 2.5 and 4.5, with a mean value $R_V = 3.5$.
\item The Galactic extinction law shows significant cluster-to-cluster variation, demonstrating that a single universal extinction curve cannot describe all Galactic environments.
\item Ultraviolet–optical colour-excess ratios provide strong constraints on extinction behaviour and are particularly sensitive to dust-grain properties.
\item The homogeneous analysis of 105 observed clusters provides one of the largest consistent extinction law samples based on combined ground-based ultraviolet and Gaia photometry.
\item The most common metallicity in the CMD is Z = 0.008; this is what we expected, as most of the clusters are of intermediate age, and the most common uncertainty is 0.01.
\end{enumerate}

Our finding of a higher mean Galactic extinction ratio $R_V = 3.53$ serves as a critical warning against using $R_V \approx 3.1$  in Galactic structure studies. Ignoring this assumption directly reduces systematic errors in open cluster distance and age estimates while providing a crucial empirical calibration anchor for future space-based ultraviolet missions.
\\
Our final goal is to fit the isochrone in the Colour-Magnitude Diagram (CMD) for all 105 observed clusters. To fit the isochrone in the CMD, we need age, distance, reddening, and metallicity. We have age and distance from Hunt \& Reffert, (2023) catalogue, and we have reddening from our analyses. We plot the CMD with metallicity as a free parameter and estimate the metallicity for all 105 sample clusters. 
\\
This catalogue constitutes an important reference for future investigations of Galactic dust structure and for improving extinction corrections in stellar population studies. The methodology presented here can be directly applied to upcoming ultraviolet surveys and next-generation photometric datasets to further refine our understanding of interstellar extinction.

\section*{Future Prospects} 

As part of our future plan, we will use the same processes for the Northern Hemisphere clusters that we have already observed to calculate their reddening laws and metallicities.
\\
The homogeneous extinction law catalogue presented here provides an important reference for future studies of Galactic dust structure and stellar population analysis. The methodology developed in this work demonstrates the diagnostic power of combining ultraviolet photometry with Gaia observations and establishes a framework for extending extinction law studies to larger Galactic samples.

\section*{Acknowledgements}

This work was supported by the MUNI/A/1419/2023. \\ 
Thanks to Ernst Paunzen for his valuable help.



\begin{thebibliography}{}

\bibitem[]{}
Akkaya Oralhan I., {\c{C}}akmak H., Karata{\c{s}} Y., Michel R., Bonatto C., 2024, {\em MNRAS} 531, 3


\bibitem[]{}
Alfonso J., Garc{\'\i}a-Varela A., Vieira K., 2024, {\em A\&A} 689


\bibitem[]{}
Barbary K., 2021, {\em Astrophysics Source Code Library}


\bibitem[]{}
Bressan A., Marigo P., Girardi L., Salasnich B., Dal Cero C., Rubele S., Nanni A., 2012, {\em MNRAS} 427, 1


\bibitem[]{}
Cavallo L., Spina L., Carraro G., Magrini L., Poggio E., Cantat-Gaudin T., Pasquato M., Lucatello S., Ortolani S., Schiappacasse-Ulloa J., 2024, {\em AJ} 167, 1


\bibitem[]{}
Fitzpatrick E. L., 1999, {\em pasp} 111, 755 


\bibitem[]{}
Gaia Collaboration., 2023, {\em A\&A} 674


\bibitem[]{}
Hunt E. L., Reffert S., 2023, {\em A\&A} 673


\bibitem[]{}
Legnardi M.~V., Milone A.~P., Cordoni G., Lagioia E.~P., Dondoglio E., Marino A.~F., Jang S., Mohandasan A., Ziliotto T., 2023, {\em MNRAS} 522, 1


\bibitem[]{}
Li L., Shao Z., Li Z., Fu X., 2025, {\em AJ} 170, 5 


\bibitem[]{}
Mowlavi N., Eggenberger P., Meynet G., Ekstr{\"o}m S., Georgy C., Maeder A., Charbonnel C., Eyer L., 2012, {\em A\&A} 541


\bibitem[]{}
Netopil M., Akkaya Oralhan I., {\c{C}}akmak H., Michel R., Karata{\textcommabelow s} Y., 2022, {\em MNRAS} 509, 1 


\bibitem[]{}
Ramezani T., Paunzen E., Gorodilov A., Pintado O.~I., 2025, {\em CAOSP} 55, 1


\bibitem[]{}
Valdes F. G., Campusano L. E., Velasquez J. D., Stetson P. B., 1995, {\em pasp} 107


\bibitem[]{}
Viscasillas V. C., Magrini L., Miret-Roig N., Wright N.~J., Alves J., Spina L., Church R.~P., Tautvai{\v{s}}ien{\.{e}} G., Randich S., 2024, {\em A\&A} 689





\end{thebibliography}
\end{document}